\documentstyle[11pt,newpasp,twoside, epsf]{article}
\def\edcomment#1{\iffalse\marginpar{\raggedright\sl#1\/}\else\relax\fi}
\reversemarginpar

\begin{document}
\title{The {\sl MAP} Satellite Mission to Map the CMB 
Anisotropy }
 \author{Lyman Page}
\affil{Princeton University, Dept. of Physics,
Jadwin Hall, Washington Rd}

\begin{abstract}
The Microwave Anisotropy Probe ({\sl MAP}) satellite is scheduled to launch in
mid-2001. {\sl MAP}'s goal is to produce a map of the 
anisotropy in the cosmic microwave background of unprecedented {\it accuracy} 
and precision. The guiding design principle has been the minimization
of systematic effects. The instrument design and mapping strategy work in
concert to take advantage of the unique opportunities afforded 
by deep space. We give an overview of the mission and compare 
the projected {\sl MAP} error bars to recent measurements.
\end{abstract}

\section{Introduction}
The cosmic microwave background (CMB) is now widely 
recognized as one of
the premier probes of cosmology. Other than foreground emission, which
is subdominant and measurable, very little astrophysics stands between the
observations and that which is of cosmological import (Tegmark et al. 2000). 
The theoretical framework is in place; the challenge before us
is performing accurate and unassailable measurements of the properties
of the CMB.

Since the discovery of the anisotropy by COBE/DMR (Smoot et al. 1992)
experimental progress has been rapid.
By the end of the last millennium, we knew that not only was there a peak
in the angular spectrum, but we knew its amplitude, width, and 
position (Hu 2000, Dodelson \& Knox 2000, Knox \& Page 2000). 
The recent BOOMERanG (de
Bernardis et al. 2000) and MAXIMA (Hanany et al. 2000)
results have not only given us maps of the CMB but have
impressively shown that peak in sharp relief. Analyses of the data
(Bond et al. 2000, Jaffe et al. 2000) give us a glimpse of what may be learned
from the primary anisotropy ($l<2000$) in the context of adiabatic
CDM models.

The future of CMB measurements is bright. In addition to results from
many ground and balloon based experiments, the {\sl MAP} satellite will 
map the CMB over the full sky to unprecedented accuracy and precision.  
{\sl PLANCK} will follow later in the decade. 
The detection and characterization of the polarization will be crucial to
verifying the picture unveiled by the temperature anisotropy and to
placing further constraints on models. At $l>2000$, the CMB
can directly probe the formation of structures through the
Ostriker-Vishniac effect, the Sunyaev-Zel'dovich effect, and lensing.
Foreground emission at these scales will be more difficult
to subtract from the data (Toffolatti et al. 1998) however.

\section{MAP in comparison with other maps}
The quest to make a large area map of the CMB anisotropy 
has been in cosmologists minds for years. Early attempts 
(reviewed by Weiss 1980 and Partridge 1995) measured the 
CMB dipole, set limits on the anisotropy, and introduced mapping
and analysis techniques. The goal
was to produce a map with a temperature and error bar for each sky pixel.
All of these experiments were done from balloons
so that atmospheric fluctuations would not skew the map.

The modern era of CMB studies started with
the $7^\circ$ resolution full-sky COBE/DMR map (Smoot et al. 1992, 
Bennett et al. 1996) which unambiguously detected the anisotropy.
This is still the map with the lowest systematic error; it is also
the best checked map. One of the most important aspects
of DMR is that pixel-to-pixel correlations are small, 
$\Sigma_{ij}=\sigma_i^2\delta_{ij}$ (where $\Sigma_{ij}$ is the
covariance between pixels $i$ and $j$) is an excellent
approximation. Lineweaver et al. (1994) showed that for $60^{\circ}$
lags, the DMR beam separation angle, the cross correlation was only
0.45\% of the diagonal terms. In the final map, the S/N per beam
resolution element was roughly two.

Next came FIRS (Meyer et al. 1991, Ganga et al. 1993) at 170 GHz 
and $3.8^{\circ}$ resolution. For FIRS
$\Sigma_{ij}\ne\sigma_i^2\delta_{ij}$ but the off diagonal
terms were small enough for the analyses. A spot check, requested 
by John Mather before the publication of Ganga et al. (1993), showed 
the largest to be of order 0.1 the diagonal elements.
At the time, there was no way to handle the full covariance matrix for
FIRS's 3500 pixels. The S/N per DMR beam was roughly $\sqrt{2}$.

QMAP (Devlin et al. 1999, Herbig et al. 1999, de Oliveira-Costa et
al. 1999), with a resolution of $0.8^{\circ}$ at 40 GHz,
came after FIRS. $\Sigma_{ij}$ was not diagonal 
for QMAP but the full covariance matrix was taken into account in the 
analysis. QMAP used the rotation around the NCP to achieve
an interlocking scan strategy. 
The features ones sees in QMAP are hot and cold spots
in the CMB. The S/N is roughly 2 per beam 
over 530 square degrees.

Most recently, we have seen the very high 
S/N ($>5$ per beam) BOOMERanG and MAXIMA 
results at $\approx 0.2^{\circ}$ resolution. Here too, the covariance 
matrix is not diagonal but has been taken into account. With their high
resolution, one can clearly see the angular scale of an acoustic
peak in the CMB. The analyzed BOOMERanG data covered 440 square degrees 
and MAXIMA covered 124 square degrees. Both experiments  
have more data which will be analyzed over the following few years.

As the sophistication in data analysis grows, more and more data sets 
are being presented as a ``map'' plus covariance matrix, so the
distinction made above will be blurred. Indeed, maps 
are now commonly synthesized from
interferometer data and scanning beam strategies. However, the 
ideal map is like DMR's: a set of temperatures and statistical weights, 
with ignorable off diagonal elements in the covariance matrix.

Producing a full-sky map with diagonal $\Sigma_{ij}$ was 
one of {\sl MAP}'s design goals. The cornerstones of achieving this
are 1) low 1/f noise in the instrument and detectors, and 2) 
a fully interlocking scan strategy\footnote{The notion of the 
interconnectedness of the scans was stressed in the proposal in 1995
but had to wait to be quantified by Wright (1996) and Tegmark (1997). }.
Concomitant with this goal we aimed to ensure that orbit and spacecraft induced
systematic errors would be negligible compared to the astrophysical signal.

\section{Instrument description}
{\sl MAP} was proposed in June 1995, at the height of the ``faster,
better, cheaper'' era; building began in June 1996. It was designed to be 
robust, thermally and mechanically stable, built of components
with space heritage\footnote{Other than the HEMT amplifiers, which were
custom designed, all components were ``off-the-shelf.'' I think it 
was a surprise
to those of us not familar with the satellite business how much testing
was required of, and how many problems there were with, 
off-the-shelf hardware.}, and relatively easy to integrate and test.

Figure 1 shows a line drawing of the {\sl MAP} satellite, though it
could be a photograph. The satellite is completely built and put together.
The long and demanding process of quality assessment is underway.
As of this writing, {\sl MAP} is well into its final observatory
level environmental test program.

As a thumbnail sketch, the {\sl MAP} instrument is comprised of ten 
symmetric, passively cooled, dual polarization, differential,
microwave receivers. There are four receivers in W-band (94 GHz),
two in V-band (61 GHz), two in Q-band (41 GHz), one in Ka band (33 GHz),
and one in K band (23 GHz). The receivers are fed by back-to-back
Gregorian telescopes. A more detailed description follows
and more information may be obtained from http://map.gsfc.nasa.gov/.

In Figure 1, the large circular structure at the bottom
is comprised of solar panels and flexible aluminized mylar/kapton 
insulation. It shields the instrument from thermal emission the Sun, Earth, 
and Moon. At launch, this is folded up so that the S/C can fit into the rocket 
fairing, though it deploys roughly 90 minutes after liftoff.

A hexagonal structure, ``hex hub,'' above the solar 
panel array holds the power supplies, instrument electronics, and 
attitude control systems. A one meter
diameter thermally insulating gamma alumina cylinder (GAC) separates the 
hex hub and the instrument. The GAC supports a 190~K thermal gradient. 

Two large (5.6 m$^2$ net) and symmetric radiators
passively cool the input optics and front-end microwave electronics
to less than 100~K. One can just make out the heat straps that connect 
the base of the radiators to the microwave components housed below the 
primary reflectors. There are no cryogens or mechanical refrigerators.

\begin{figure}
\plotone{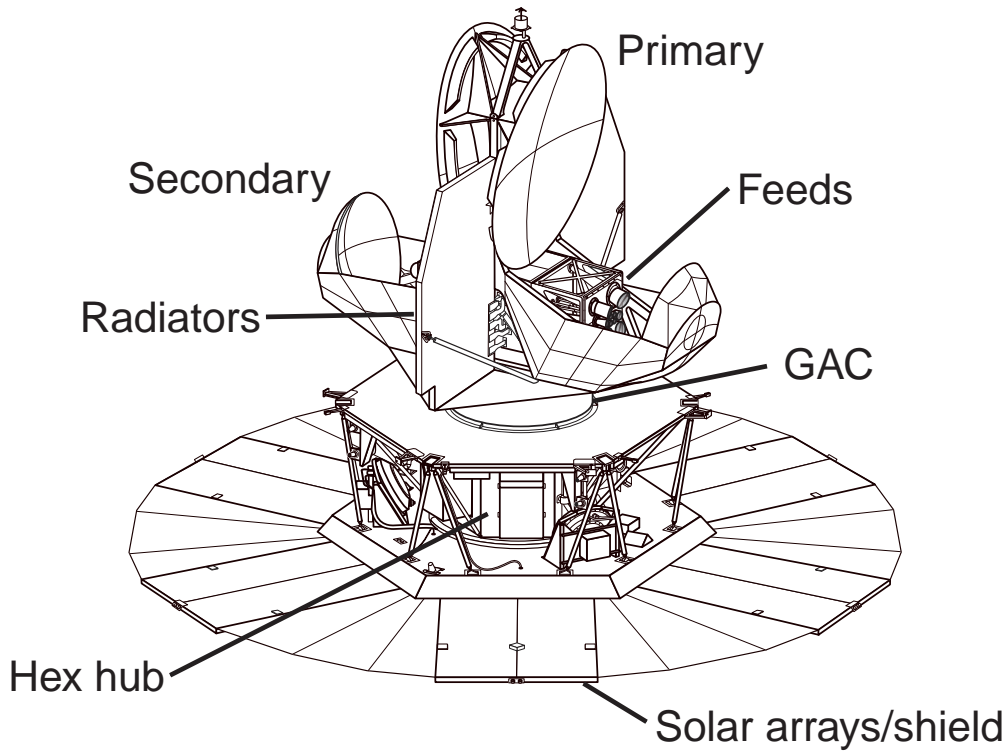}
\caption{Outline of the {\sl MAP} satellite. The overall height is
3.6~m, the mass is 830 kg, and the diameter of the large disk on 
the bottom is 5.1 m. Six solar arrays on the bottom of this disk 
supply the 400 Watts to power the spacecraft and instrument.
Thermal blanketing between the hex hub and GAC, and
between the GAC and radiators, shield the instrument from thermal
radiation from the support electronics and attitude control systems.
{\sl MAP} will be launched in mid-2001 by a Delta 7425-10 from Kennedy 
Space Center. The mission life is 27 months.}
\end{figure}

\subsection{Optics} The optics comprise two back-to-back shaped
Gregorian telescopes. The primary mirrors are 1.4 by 1.6 m. The
secondaries are roughly a meter across though most of the surface 
simply acts as a shield to prevent the feeds from directly viewing the 
Galaxy. The telescopes illuminate ten scalar feeds on each side, 
a few of which are visible in the figure. 
The primary optical axes are separated by $141^{\circ}$ to allow
differential measurements over large angles on a fast time scale. 
The feed centers occupy a 18 by 20 cm region in the focal plane,
corresponding to a $4^{\circ}$ by $4.5^{\circ}$ array on 
the sky.

At the base of each feed is an orthomode transducer (OMT) that sends the 
two polarizations supported by the feed to separate receiver chains.
The microwave plumbing is such that a single receiver chain 
(half of a ``differencing assembly'') differences electric fields with
two nearly parallel linear polarization vectors, one from each telescope.

Precise knowledge of the beams is essential for accurately computing the 
CMB angular spectrum and for calibration. Because of the large focal
plane the beams 
are not symmetric, as shown in Table 1. Cool down distortions
of the optics will alter the W-band beams with respect to the 
ground based measurements so we give only an upper bound at this time.
All beam profiles will be mapped in flight with Jupiter and 
other celestial sources.

\begin{table}
\begin{center}
\caption{Approximate Instrument Characteristics by Frequency Bands}
\begin{tabular}{cccccc} \hline
Band & $f_{center}$ (GHz) & $\Delta f_{noise}$ (GHz) & $T_{HEMT}$ (K) 
     & $N_{chan}$ & $\theta_{FWHM}$ (deg)\\
K  & 23 & 5 & 25 &2 & $0.75^{\circ}~by~0.95^{\circ}$    \\
Ka & 33 & 7 & 35 &2 &   $0.6^{\circ}~by~0.7^{\circ}$    \\
Q  & 41 & 8 & 50 &4 &    $0.45^{\circ}~by~0.5^{\circ}$    \\
V  & 61 & 11 & 80 &4 &   $0.3^{\circ}~by~0.35^{\circ}$    \\
W  & 94 & 18 & 100 &8 & $<0.23^{\circ}$ \\
\end{tabular}
\end{center}
\end{table}

One of the design constraints was to minimize stray radiation from the 
Galaxy, Sun, and Earth. We use physical optics codes
to compute the sidelobe pattern over the full sky as well
as to compute the current distributions on the optics. We have also built a
specialized test range to make sure that, by measurement, we can 
eliminate the Sun as a source of signal at $<1~\mu$K level in all
bands. This requires knowing beam profiles down to 
roughly $-45$ dBi (gain above 
isotropic) or $-105$ dB from the W-band peak. We find that over much of
the sky, the measured profiles differ from the predictions at the $-50$~dB
level due to scattering off the feed horns and the structure that 
holds them. 

We use a combination of the models and measurements to place a limit on 
the Galactic pickup in the sidelobes. At 90 GHz, less than 
$2~\mu$K of Galactic signal should contaminate the boresite signal for
observations with $|b|>15^{\circ}$, before any modeling. 
As this adds in quadrature to the CMB signal, the affect on the
angular spectrum will be negligible. The characterization of the 
sidelobes has been done by Chris Barnes.

\subsection{Receivers}
{\sl MAP} uses ``pseudo-correlation'' receivers (Jarosik 2000) to 
measure the difference in
power coming from the outputs of the OMTs at the base of the feeds,
as shown in Figure 2. We use 
the term ``pseudo'' to refer to the fact that these radiometers use
two hybrid tees and two square law detectors in place of the multiplier
in a true correlation receiver.
The ``correlation'' refers to the feature that the system 
is primarily sensitive to the correlated signal in the two arms.

The {\sl MAP} mission was made possible by the HEMT-based amplifiers developed
by Marian Pospieszalski (1992) at NRAO.  These amplifiers achieve noise
temperatures of 25-100 K at 80~K physical temperature (Pospieszalski \&
Wollack 2000). Of equal importance is that the amplifiers can be phase matched 
over a 20\% fractional bandwidth, as is required by the receiver design.

\begin{figure}
\plotfiddle{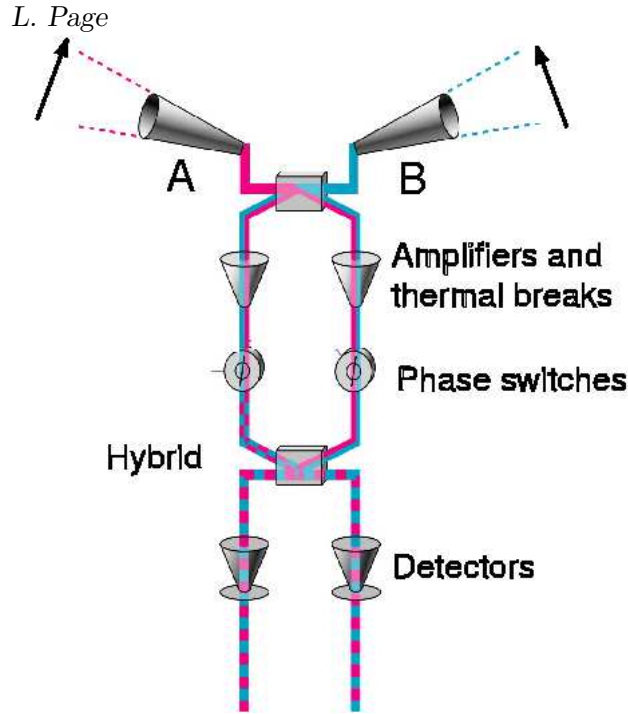}{3.125truein}{0}{45}{45}{-110}{-10}
\caption{One half of a differencing assembly for detecting
one polarization component. Hybrids (``magic tees'') split the
inputs into two arms where the signal is amplified before recombining.
There are actually two stages of amplification. The microwave filters 
between the lower hybrid and the detectors are not shown.}
\end{figure}

In Figure 2, radiation from the two
feeds is combined by a hybrid into $(A+B)/\sqrt{2}$ 
and $(A-B)/\sqrt{2}$ signals,
where A and B refer to the amplitudes of the electric fields from one
linear polarization of each feed horn. In one arm of the
receiver, both A and B signals are amplified first by cold ($<100~$K) and
then by warm (290 K) HEMT amplifiers. Noise power from the amplifiers,
which far exceeds the input power, is added to each signal by the first 
amplifiers. Ignoring the phase switch for 
a moment, the two arms are then recombined in a second hybrid and both 
outputs of the hybrid are detected after a band defining filter.
Thus for each differencing assembly,
there are four detector outputs, two for each polarization.
In a perfectly balanced system, one
detector continuously measures the power in the A signal plus the average
radiometer noise while the other continuously measures the power in the
B signal plus the average radiometer noise.

If the amplification factors for the fields are $G_1$ and $G_2$ for the two
arms, the difference in detector outputs is $G_1G_2(A^2-B^2)$. Note that
the average power signal present in both detectors has cancelled so that
small gain variations, which have a $1/f$ spectrum, act on the 
difference in powers from the two arms, which corresponds to less than 
1~K, rather than on the total power which corresponds to roughly 100~K in
W-band. The system is stabilized further by toggling one of the phase switches
at 2.5 kHz and coherently demodulating the detector outputs. 
(The phase switch in the other arm is required to preserve the phase
match between arms and is jammed in one state.) The 2.5 kHz modulation
places the desired signal at a frequency above the $1/f$ knee
of the detectors and video amplifiers as well as rejecting any residual
effects due to $1/f$ fluctuations in the gain of the HEMT amplifiers.
The power output of
each detector is averaged for between 51 and 128 ms and telemetered
to the Earth. In total, there are forty signals (only half
contain independent information) plus instrument housekeeping data 
resulting in a data rate of 110~MBy/day.

%
%

The power spectrum of the
noise shows that it is effectively white between 0.008 Hz, 
the spin rate of the 
satellite, and 2.5~kHz. The autocorrelation function shows only a spike
at zero lag and a 1.2\% correlation between adjacent samples (in W-band)
due to the antialiasing filter.
All tests show that the noise is stationary and Gaussian
for days at a time.

Although {\sl MAP}'s differential design was driven by the $1/f$ noise in
the amplifiers, it is also very effective at reducing the effects of
$1/f$ thermal 
fluctuations of the spacecraft itself. The thermal stability of deep space
combined with the insensitivity to the spacecraft's
slow temperature variations should result in an extremely stable 
instrument. Outside of the antialiasing filter and finite sampling time,
effects which are computable, measured, and non-random, we have not been
able to identify other effects that will correlate one measurement to the next.

%
%
%

\subsection{Scan strategy}
The other key aspect of producing a map with 
$\Sigma_{ij}=\sigma_i^2\delta_{ij}$, in addition to low system $1/f$,
is a highly interlocking scan 
strategy. In any measurement, a baseline instrumental offset along with
its associated drift, must be subtracted. Without cross hatched scans 
this subtraction can correlate pixels over large swaths, 
resulting in striped maps and substantially more involved 
analyses.

{\sl MAP} observes from a low maintenance Lissajous orbit at L2, with the 
Sun Earth and Moon always behind, as shown in Figure 3. 
Corrections to the orbit are applied roughly once per
season through thruster jets; there is only one mode of operation. 
{\sl MAP} will be the first satellite to stay at L2
for an appreciable time.
%

{\sl MAP} spins around its axis with a period of 2 min
and precesses around a $22.5^{\circ}$ degree cone every hour so that the beams
follow a spirograph pattern. Consequently, $\approx 30$\% of the sky 
is covered in one hour, before the instrument temperature 
can change appreciably. This motion is accomplished with three
spinning momentum wheels; the net angular momentum of the satellite is
near zero. 
The axis of this combined rotation/precession sweeps out approximately a 
great circle as the Earth orbits the sun. In six months, the whole sky
is mapped. Systematic effects at the spin 
period of the satellite are the most difficult to separate from true 
sky signal. Such effects, driven by the Sun, are minimized
because the instrument is always in the 
shadow of the solar array and the precession axis is fixed with
respect to the Earth-Sun line.

\begin{figure}
\plotone{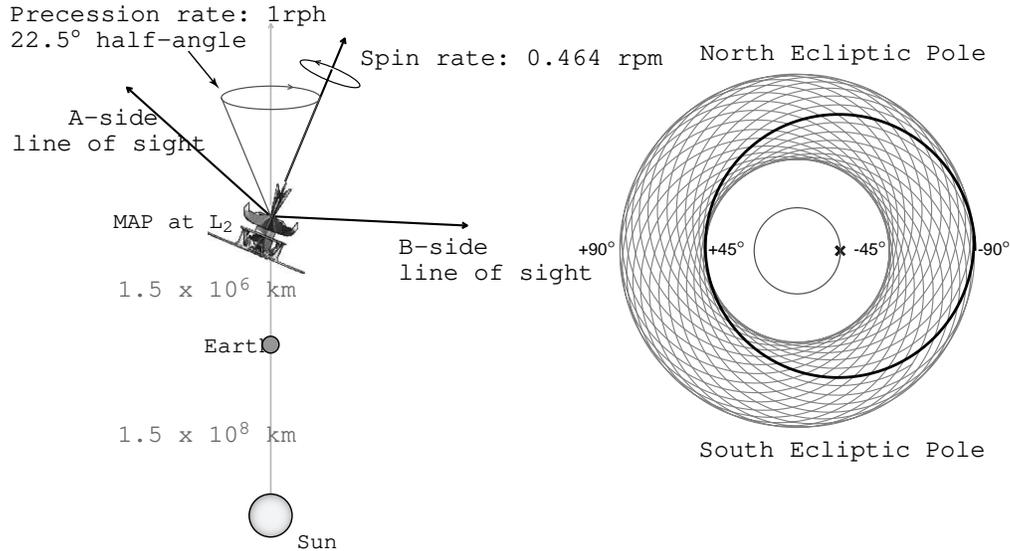}
\caption{{\sl MAP's} scan pattern from L2. The dark circle on the lefthand
drawing depicts the path covered by two beams for one rotation, the
innermost circle is the path of the spin axis during one precession.}
\end{figure}

The combination of {\sl MAP}'s four observing time scales (2.5 kHz, 2.1 min,
1 hour, 6 months) and the heavily interlocked pattern result is a strong
spatio-temporal filter for any signal fixed in the sky. A pipeline
for simulating the mission, the amplifier characteristics, the beam
profiles, and producing maps from the time ordered 
data is written. Using the measured amplifier
characteristics, we can form the two point function of a simulated noise
map. It shows that the pixel-pixel correlations are negligible
with the exception of a $<1\%$ nearest neighbor correlation from the
antialiasing filter and a $\sim0.1\%$ correlation at angles of 
the beam separation. See Hinshaw (2000) for more details.

\subsection{Science from {\sl MAP}}
{\it We emphasize again that the primary goal of {\sl MAP} 
is to produce high
fidelity polarization-sensitive full-sky multi-frequency maps of the microwave
sky.} With such maps we can not only determine the anisotropy in the
CMB but we can test our fundamental assumptions about the cosmos.
For instance, perhaps the best fit model is not a simple adiabatic CDM variant 
but has some isocurvature modes mixed in. 

If the CMB temperature fluctuations are Gaussian with random phases,
as the current data suggest, then the cosmological information
may be obtained from the angular spectrum. An example is shown in Figure
4. The limit to which we can
know the angular spectrum is set by the number of independent patches of sky
corresponding to a given angular scale, or equivalently the number
parameters needed to describe a mode. For instance, five parameters
determine the quadrupole. No matter how well we measure these,
we cannot know the fractional variance of the parent distribution
better than $\sqrt{2/(2l+1)}=0.63$. This is
the ``cosmic variance'' limit and is one of the motivations for a
full-sky map. {\sl MAP} will be cosmic variance limited
up to $l\approx 500$ in W-band. In other words, in principle it is 
not possible to
determine the angular spectrum better than this. More information
may be obtained from high S/N polarization measurements however. 

One of the largest uncertainties of existing measurements is 
the calibration. Limitations come from not knowing the beam profiles
precisely and from the intrinsic uncertainty in the calibration sources.
The flux from planets is known to $\approx 5$\%
depending on frequency. Although the dipole has been used to calibrate 
a number of mapping experiments (Meyer et al. 1991, de Bernardis et al. 2000, 
Hanany et al. 2000), it is difficult to measure with limited sky coverage
because of its covariance with typical scan patterns. {\sl MAP} like 
DMR, will calibrate using the change in the dipole caused by the
Earth's motion around the Sun. The calibration error will be $<1$\%. 

In Figure 4, we compare the projected {\sl MAP} uncertainties with the 
state-of-the-art. We have included calibration error in both plots.
The right hand plot shows the
amplitude and position of the first peak ($50<l<420$) based on 
Knox \& Page (2000). This is a relatively {\it cosmological 
model independent} parametrization of the peak. It differs from the 
work of Bond and colleagues (see these proceedings) in that a 
CDM variant is not assumed and there
is no marginalization over nonrelevant parameters, consequently the
error bars are smaller. Though 
one should not put too much stock in interpreting the contours
beyond the $\approx 2\sigma$ level, the plot shows that, despite high 
precision, we are still not in complete agreement on the position and 
amplitude of the first peak. 

There will be much more to do with the {\sl MAP} data than fit cosmological
models. A partial list includes understanding Galactic 
foreground emission (is there spinning dust?) and extragalactic sources
(they are separated from the CMB through their frequency spectrum and
positive definiteness),
a search for time variable sources, a search for non-standard 
topologies, a measurement of the polarization-temperature cross 
correlation, a search for the large scale 
Sunyaev-Zel'dovich effect, a measurement of the 
cross-correlation with the Solan Digital
Sky Survey to assess structure formation at lower redshift ($z<1000$).

Over the next two years, our current picture of the primary anisotropy
will become clearer
as more BOOMERanG and MAXIMA data are analyzed and as new data come in
from ACBAR, Arkeops, BEAST, CBI, DASI, MINT, TopHat, VSA, and others. The full
data set will of course be interesting on its own, and will
be essential for ensuring that the picture we get from {\sl MAP} extrapolates
as predicted in both frequency and angular scale. The next generation
of polarization measurements of the primary anisotropy will take
us beyond {\sl MAP}'s polarization sensitivity and add a new 
dimension to CMB studies. It will not be 
too long before CMB experiments will be calibrated on the primary
anisotropy, a prospect difficult to believe just a few years ago.

\begin{figure}
\plottwo{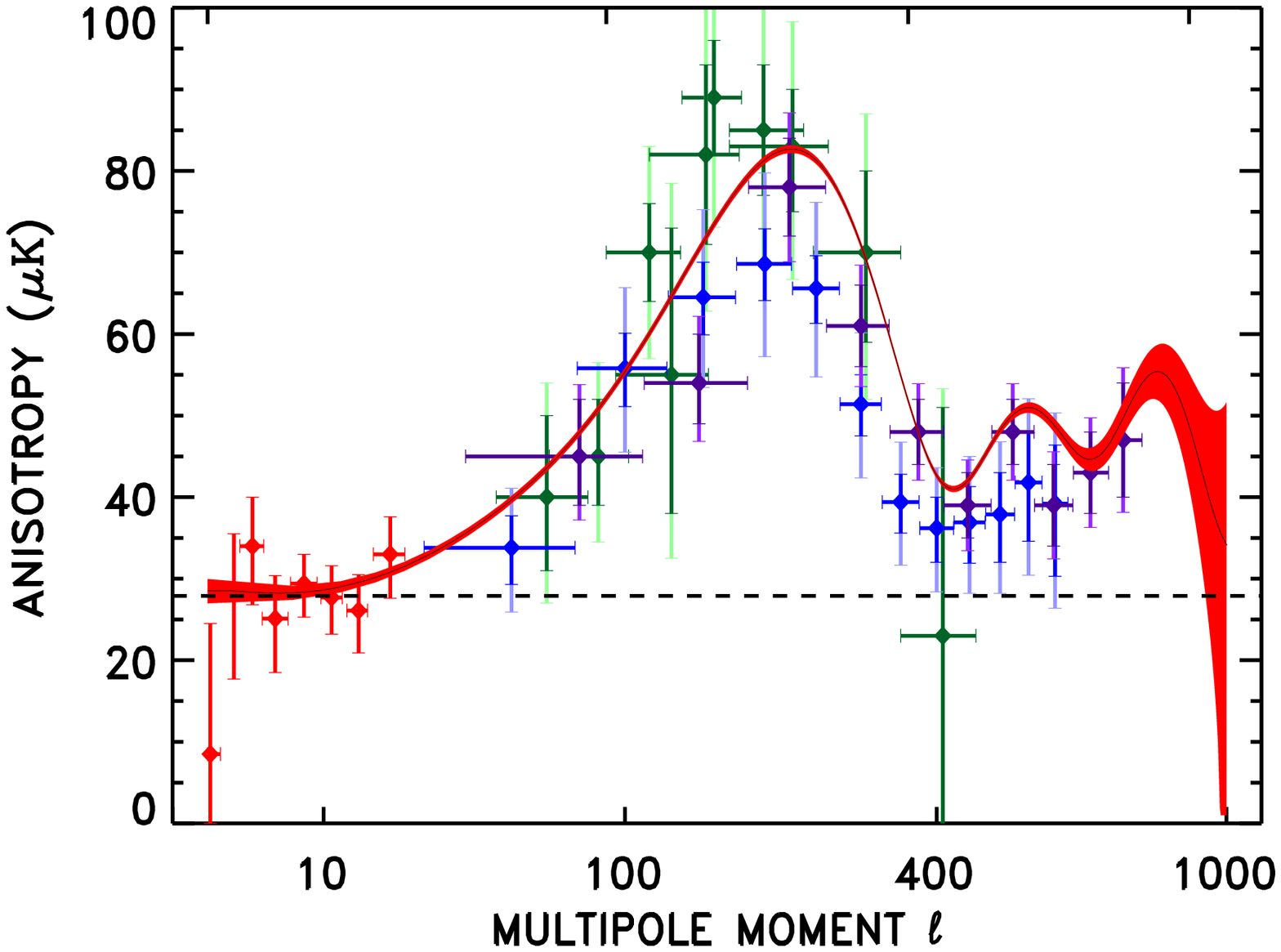}{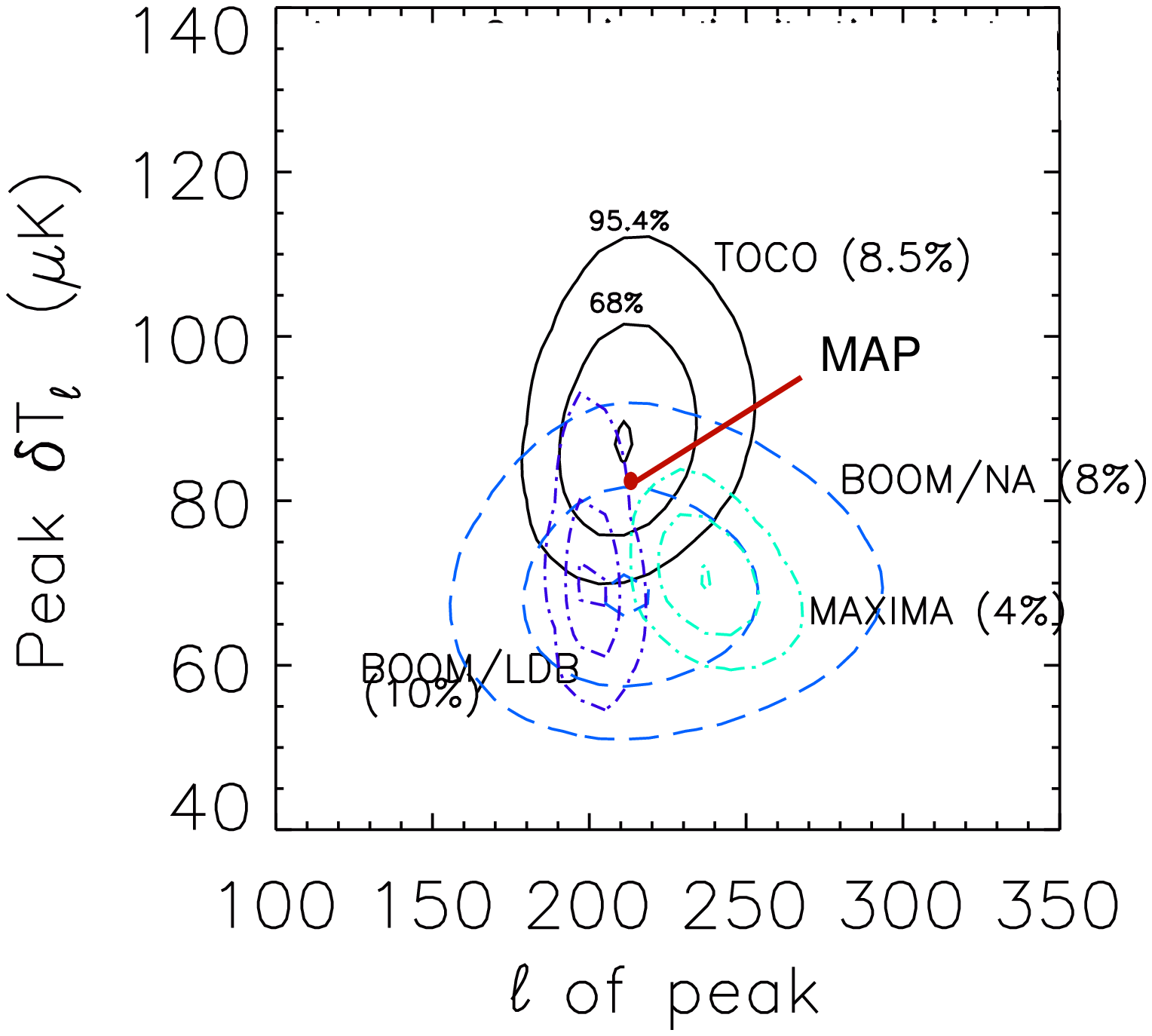}
\caption{{\it Left.} The projected {\sl MAP} error bars for a band
averaging of $\Delta l=50$ superimposed on a popular model
along with the BOOMERanG, MAXIMA, and TOCO
(Torbet et al. 1999, Miller et al. 1999) data. The calibration
uncertainty is indicated by the light lines behind the thick lines.
It is not usually plotted like this because the calibration is 
common to all data in a set. {\it Right.} An analysis of the peak
following the Knox \& Page (2000) Gaussian/Temperature method.
Calibration error is included and shown in parentheses. The BOOMERanG 
N/A data are from Mauskopf et al. (2000). The small solid dot shows the
projected {\sl MAP} error.}
\end{figure}

\acknowledgements
The work on which this article is based, except for the righthand panel
in Figure 4,
was done by the {\sl MAP}
science team\footnote{Chuck Bennett (NASA/GSFC, PI),
Mark Halpern (UBC), Gary Hinshaw (NASA/GSFC), Norm Jarosik (Princeton),
Al Kogut (NASA/GSFC), Michele Limon (Princeton), Stephan Meyer
(Chicago), Lyman Page (Princeton), David Spergel (Princeton), Greg
Tucker (Brown), David Wilkinson (Princeton), Ed Wollack (NASA/GSFC),
and Ned Wright (UCLA).} and the {\sl MAP} satellite project led 
by Liz Citrin (Project Manager) and Cliff Jackson (Systems Engineer)
at NASA/GSFC. Over a hundred NASA employees and contractors have 
dedicated themselves to making {\sl MAP} work.

%
%
%
%

\end{document}